\documentclass[5p,number,sort&compress]{elsarticle}
\usepackage{graphicx}
\usepackage{amsmath,amssymb}
\usepackage[breaklinks,colorlinks=true]{hyperref}
\usepackage{slashed}
\usepackage{mathtools}
\usepackage{braket}
\usepackage{simplewick}
\allowdisplaybreaks

\newcommand{\Nc}{N_\text{c}}

\newcommand{\mui}{\mu_i}
\newcommand{\id}{\text{id}}
\newcommand{\muhat}{\hat{\mu}}
\newcommand{\bmu}{\boldsymbol{\mu}}
\newcommand{\MeV}{\,\text{MeV}}
\newcommand{\fm}{\,\text{fm}}
\newcommand{\kFi}{k_{\mathrm{F},i}}
\newcommand{\Bbar}{\bar{B}}

\newcommand{\thermal}{\text{th}}

\newcommand{\nB}{n_{B}}
\newcommand{\muB}{\mu_{B}}

\begin{document}

\title{
  \vspace*{-2em} {\small \hfill KEK-TH-2347, J-PARC-TH-0250, RIKEN-iTHEMS-Report-21} \\
  \vspace{1.5em}  Equation of state of neutron star matter and its warm
  extension \\ with an interacting hadron resonance gas}

\author[UT]{Yuki~Fujimoto}
\ead{fujimoto@nt.phys.s.u-tokyo.ac.jp}
\author[UT]{Kenji~Fukushima}
\ead{fuku@nt.phys.s.u-tokyo.ac.jp}
\author[KEK,Sokendai,Riken]{Yoshimasa~Hidaka}
\ead{hidaka@post.kek.jp}
\author[NYCU,Kochi]{Atsuki~Hiraguchi}
\ead{a.hiraguchi@nycu.edu.tw}
\author[Kochi]{Kei~Iida}
\ead{iida@kochi-u.ac.jp}

\address[UT]{Department of Physics, The University of Tokyo, %
  7-3-1 Hongo, Bunkyo-ku, Tokyo 113-0033, Japan}
\address[KEK]{Institute of Particle and Nuclear Studies, KEK, %
1-1 Oho, Tsukuba, Ibaraki 305-0801 Japan}
\address[Sokendai]{Graduate University for Advanced Studies (Sokendai), Tsukuba 305-0801, Japan}
\address[Riken]{RIKEN iTHEMS, RIKEN, 2-1 Hirosawa, Wako, Saitama 351-0198, Japan}
\address[NYCU]{Institute of Physics, National Yang Ming Chiao Tung University, Hsinchu 30010, Taiwan}
\address[Kochi]{Department of Mathematics and Physics, %
  Kochi University, Kochi 780-8520, Japan}

\begin{abstract}
  We propose an interpolating equation of state that satisfies
  phenomenologically established boundary conditions in two extreme
  regimes at high temperature and low baryon density and at low
  temperature and high baryon density.
  We confirm that the hadron resonance gas model with the
  Carnahan-Starling excluded volume effect can reasonably fit the
  empirical equation of state at high density up to several times the
  normal nuclear density.  We identify the onsets of strange particles
  and quantify the strangeness contents in dense matter.  We finally
  discuss the finite temperature effects and estimate the thermal
  index $\Gamma_{\rm th}$ as a function of the baryon density, which
  should be a crucial input for the core-collapse supernova and the
  binary neutron star merger simulations.
\end{abstract}
\maketitle

\section{Introduction}
\label{sec:intro}

Strongly interacting matter has been investigated well in two extremes
of the temperature $T$ and baryochemical potential $\muB$.
The heavy-ion collision experiments realize matter at high-$T$ and
low-$\muB$, which can also be studied by the Monte-Carlo simulation of
quantum chromodynamics (QCD) on the
lattice~\cite{Borsanyi:2013bia, Bazavov:2014pvz}.
Another extreme matter at low-$T$ and high-$\muB$ is found in cores of
neutron stars, for which the Monte-Carlo lattice-QCD simulation is of
no use due to the notorious sign problem.

Confronting the lattice-QCD thermodynamics, the hadron resonance
gas (HRG) model has become a common tool to approximate the equation
of state (EoS) below the critical temperature $T_c \sim 150\MeV$
(see, e.g., Ref.~\cite{Andronic:2017pug} for a recent review).
The HRG model is a parameter-free description of thermal properties
with experimentally observed particle and resonance spectra.
In the HRG model, nonperturbative hadronic interactions are assumed to
be dominated by poles and branch cuts corresponding to
resonances~\cite{Hagedorn:1967tlw}.
The ideal HRG (IHRG) model consists of noninteracting hadrons and
resonances, and yet, its validity has been endorsed by the successful
thermal fit of particle abundances in heavy-ion collision experiments.
The IHRG EoS is recently extended to the terrain of finite
baryochemical potential in a region of
$\mu_B / 3 \lesssim T$~\cite{Monnai:2019hkn}, guided by the conserved
charge susceptibilities from lattice-QCD
calculations~\cite{HotQCD:2012fhj, Ding:2015fca, Bazavov:2017dus}.
This region may include a possibility of Quarkyonic
matter~\cite{McLerran:2007qj, Kojo:2009ha, Kojo:2011cn,
  Lottini:2012as, Philipsen:2019qqm} and the associated triple-point
structure in the QCD phase diagram~\cite{Andronic:2009gj}.
Furthermore, owing to its parameter-free nature, it can also be
applied to a wide variety of problems such as a reference for baryon
number fluctuations~\cite{Fukushima:2014lfa}, a shift of the chemical
freezeout due to the inverse magnetic
catalysis~\cite{Fukushima:2016vix}, and the rotational effect on the
deconfinement temperature~\cite{Fujimoto:2021xix}, to mention a few.

Thermodynamic quantities from the IHRG model, however, blow up above
$T_c$.  We expect that the validity region of the HRG model could be
extended to higher temperature by introducing the
\textit{interaction effect}.
The excluded volume (EV) effect is the simplest way to implement the
interacting HRG and the formulation was given in
Ref.~\cite{Andronic:2012ut};  the short-range repulsive interaction
was modeled via a hard-core correction following the thermodynamically
consistent way as developed in Ref.~\cite{Rischke:1991ke} (see also
Ref.~\cite{Hagedorn:1980kb}).  By supplementing the repulsive
interaction with an additional attractive interaction, the interacting
HRG model amounts to the van der Waals (VDW)
EoS~\cite{Vovchenko:2016rkn, Vovchenko:2017zpj} (see also
Ref.~\cite{Vovchenko:2020lju} and references therein).
Albeit a few known theoretical problems~\cite{Lo:2017ldt}, this crude
model approach is successful in describing the lattice-QCD results
including the EoS and the cumulants (see, e.g., Refs.~\cite{Andronic:2012ut,
  Vovchenko:2016rkn}). Indeed the nuclear interaction has an intricate
structure, but the nuclear matter properties can be captured by two
parameters (see, e.g., Refs.~\cite{PhysRevA.97.013601, Ohashi:2020djc}
for the two-parameter nuclear matter description by the scattering
length and the effective range).

Up to here, we have reviewed the interacting HRG model in the context of
high-$T$ and low-$\muB$ studies.  Then, it is a natural anticipation
that the same machinery of the interacting HRG model can also work at
low-$T$ for astrophysical applications;  interestingly enough, this
idea of applying the HRG description for neutron stars can be traced
back to the pioneering work by Hagedorn~\cite{Hagedorn:1965st} more
than half a century ago even before the discovery of pulsars.
One might think that this anticipation is immediately falsified by the
nuclear liquid-gas phase transition of first order in the low-$T$ and
high-$\muB$ region.  Although the IHRG model is unable to describe the
first-order phase transition, the interacting VDW-HRG model can
properly account for the liquid-gas phase transition as studied in
Refs.~\cite{Vovchenko:2016rkn, Vovchenko:2015xja}.

In the present work, we will look into the EoS from the VDW-HRG model
by making a quantitative comparison with the Chiral Effective Field
Theory ($\chi$EFT).  We will show that the hard-core EV effect
violates the causality, but the VDW-HRG model with refined repulsive
interaction leads to more reasonable behavior at low $T$ and high $\muB$.
The modified repulsive interaction is incorporated \`a la
Carnahan-Starling (CS)~\cite{Carnahan:1969}.  The CS excluded volume
has been adopted in the hadron physics~\cite{Anchishkin:2014hfa,
  Satarov:2014voa, Satarov:2016peb}, and
astrophysics~\cite{Lourenco:2019ist, Dutra:2020qsn}.  Within the HRG
model with the CS-type EV, i.e., the CS-HRG model, we will construct
the EoS, $p(\nB)$ (the pressure vs.\ the baryon number density), for
dense and warm matter at $T < 60\MeV$.
Even though the parameters in the CS-HRG model are fixed by the
$\chi$EFT, the CS-HRG can provide us with useful insights.  One
example is that we can diagnose the strangeness contents in dense
matter hadron by hadron.  Another useful application is the finite-$T$
extension of the EoS;  in many simulations of binary neutron star
mergers the thermal component of the EoS is treated as an ideal gas for
convenience~\cite{Bauswein:2010dn,Figura:2020fkj}, and there is a recent discussion
in the astrophysical context~\cite{Fore:2019wib} that thermal pions
could be important.

At the conceptual level, the interacting HRG model embodies the idea of
Quarkyonic matter that claims continuous duality between baryonic and
quark matter (see, however, Ref.~\cite{Torrieri:2010gz} for the
earlier analysis based on the VDW model indicating the large-$\Nc$
transition between $\Nc=3$ and $\Nc\to\infty$).
The idea is consonant to the crossover EoS construction based on
quark-hadron continuity~\cite{Masuda:2012kf, Baym:2019iky,
  Fujimoto:2019sxg}.
There are theoretical attempts to build a Quarkyonic
model~\cite{Fukushima:2015bda, McLerran:2018hbz, Jeong:2019lhv,
  Sen:2020peq, Duarte:2020xsp, Duarte:2020kvi, Zhao:2020dvu,
  Cao:2020byn}, and as argued in a quantum percolation
picture~\cite{Fukushima:2020cmk} quark degrees of freedom emerge from
interactions.  In this sense, the interacting HRG model could be
regarded as a concrete modeling that exhibits the quark-hadron duality
from the hadronic side.  This theoretical argument also justifies the
extended validity of the interacting hadronic model for high-$T$ and/or
high-$\muB$ matter in which quark degrees of freedom could be mixed
together. 

We note that, throughout this work, we use natural units;
$c = \hbar = k_{\rm B} =1$.

\section{Interacting hadron resonance gas model}
\label{sec:hrg}

The HRG encompasses the numerous contributions from
experimentally measured states of hadrons and resonances.
The total thermodynamic quantities, such as the pressure, can be
decomposed into three pieces:
\begin{equation}
  p(T,\bmu) = p_M(T,\bmu) + p_B(T,\bmu) + p_{\Bbar}(T,\bmu)\,,
  \label{eq:pHRG}
\end{equation}
where $M$, $B$, and $\Bbar$ denote contributions from mesons, baryons,
and anti-baryons, respectively, and $\bmu = (\mu_B, \mu_Q, \mu_S)$ are
the chemical potentials conjugate to the net baryon number $B$, the
electric charge $Q$, and the strangeness $S$.
We here limit ourselves to $B=1$ baryons and $B=-1$ anti-baryons,
and we discard composite baryons with $B > 1$ such as the deuteron,
light nuclei, hypernuclei, etc.;  the deuteron, for example becomes
unbound in neutron-rich matter.  Moreover, since matter itself is a
gigantic nucleus (if it is self-bound), light cluster contributions
may lead to double-counting problems.
For our HRG model in this work, we have adopted the particle data
group list of particles and resonances (where the resonances are
handled in the zero-width approximation;  see Ref.~\cite{Lo:2017lym}
for a refined $S$-matrix treatment of resonances) contained in the
THERMUS-V3.0 package~\cite{Wheaton:2004qb}.

\subsection{Ideal hadron resonance gas (IHRG)}

In the IHRG model, each particle is treated as the Bose/Fermi ideal
(i.e., non-interacting) gas.  In this model, the mesonic, the baryonic,
and the anti-baryonic contributions in Eq.~\eqref{eq:pHRG} are given
by
\begin{equation}
  \label{eq:pBIHRG}
  p_{M/B/\bar{B}}(T, \bmu) = \sum_{i \in M/B/\bar{B}} p_i^{\id} (T, \mui) \,,
\end{equation}
where the pressure function $p_i^{\id}$ in the grand canonical
ensemble is
\begin{equation}
    p_i^{\id}(T, \mui) = \pm\frac{g_iT}{2\pi^2} \int_0^{\infty} k^2dk\ln\left[1 \pm e^{-(E_i- \mui) / T}\right]\,.\label{eq:pGC}
\end{equation}
We note that the overall $+$ ($-$) sign corresponds to the fermion
(boson).  The energy dispersion is
$E_i =\sqrt{k^2+m_i^2}$ with the spin degeneracy factor,
$g_i = 2s_i + 1$,
and the chemical potential, $\mu_i = B_i \muB + Q_i \mu_Q + S_i \mu_S$,
of the particle species $i$.

The other thermodynamic quantities such as the number density $n_i$ and the
energy density $\varepsilon_i$ can also be derived accordingly as
\begin{align}
  n_i^{\id}(T, \mui) =& \frac{g_i}{2\pi^2} \int_0^{\infty} \frac{k^2dk}{e^{(E_i - \mui)/T} \pm 1}\,,\label{eq:nGC}\\
  \varepsilon_i^{\id}(T, \mui) =& \frac{g_i}{2\pi^2} \int_0^{\infty} \frac{k^2dk}{e^{(E_i - \mui)/T} \pm 1} E_i\,.\label{eq:eGC}
\end{align}
For cold matter at $T=0$ we can analytically carry out the fermion
integrals (and mesons are irrelevant) to find the pressure,
$p_i^{\id}(\mui) = \frac{g_i}{24 \pi^2} \!\left[\mui \kFi
  \left(\mu_i^2 \!-\! \tfrac52 m_i^2\right) \!+\! \tfrac32 m_i^4
  \ln\left(\frac{\kFi \!+\! \mui}{m_i}\right)\right]$,
and the number and the energy densities are
$n_i^{\id}(\mui)  = \frac{g_i}{6\pi^2}\kFi^3$
and 
$\varepsilon_i^{\id}(\mui) = \frac{g_i}{16 \pi^2}\! \left[\mui \kFi
  (2 \mui^2 \!-\! m_i^2) \!-\! m_i^4 \ln\left(\frac{\kFi \!+\! \mui}{m_i}\right)\right]$,
with the Fermi momentum being $\kFi = \sqrt{\mui^2 - m_i^2}$.

\subsection{Van der Waals hadron resonance gas (VDW-HRG)}

In this work, we incorporate the interaction effect based on the VDW
construction, which comprises the repulsive interaction by the EV
effect and the attractive interaction.
The VDW EoS was originally formulated in classical systems in the
canonical ensemble with the fixed number of particles, and thus
the following two extensions were necessary:  the reformulation in the grand
canonical ensemble was given in Ref.~\cite{Vovchenko:2015xja}, and the
quantum statistics was taken into account in
Refs.~\cite{Vovchenko:2015vxa, Redlich:2016dpb}.
Hereafter, we will employ the formulation in
Ref.~\cite{Vovchenko:2017cbu}.
At this point, we shall make a comment on the interactions.
We will consider the interaction effect
only in the $B$ sector.  At zero and low temperatures, the $\Bbar$
sector is simply  negligible.
The reason why we neglect the interaction in the $M$ sector is that
mesons are only weakly interacting in the limit of large colors.
Indeed, the substantial mesonic EV effect leads to the discrepancy
between the EV-HRG and the lattice-QCD data due to too strong
suppression of thermodynamic quantities, which was already implied in
Figure~4 of Ref.~\cite{Andronic:2012ut}.

To introduce the repulsive interaction through the EV effect, we
replace the volume $V$ in the partition function by $f(\eta)V$ with
$\eta$ being the packing fraction\footnote{Strictly speaking, the EV effect has to be
  introduced through the canonical partition function as formulated originally.}.
The function $f(\eta) \in (0, 1]$ measures the volume fraction
in which particles with hard sphere can move around.
The concrete expression of $f(\eta)$ will be given soon in
Eqs.~(\ref{eq:fvdW}), and (\ref{eq:fCS}).
The packing fraction, $\eta$, can be related to the hard-sphere radius,
$R$, as $\eta = \tfrac{4\pi}3 R^3 n$ (see
Ref.~\cite{Vovchenko:2017cbu} for more description).

The VDW-HRG model in the $B$ sector can be obtained from the pressure
defined by
\begin{equation}
  p_B(T,\bmu) = [f(\eta_B) \!-\! \eta_B f'(\eta_B)] \sum_{i \in B}
  p_i^{\id}(T,\muhat_i) \!-\!a\nB^2\,,
  \label{eq:pvdW}
\end{equation}
where the last term $\propto a$ represents the attractive interaction
effect, and the shifted chemical potential $\muhat_i$ is
\begin{align}
  \muhat_i &= \mu_i + \Delta\muB\,, \label{eq:vdwhrg2}\\
  \Delta\muB &= \frac{b_B}4 f'(\eta_B) \sum_{i \in B} p_i^{\id}(T,\muhat_i) + 2a \nB\,,
  \label{eq:vdwhrg3}
\end{align}
with $b_B \equiv \tfrac12 \cdot \tfrac{4\pi}3 (2R)^3$ being the
eigenvolume of baryons.

From Eq.~\eqref{eq:vdwhrg3}, the energy density is immediately derived
as
\begin{equation}
  \varepsilon_B(T,\bmu) = f(\eta_B) \sum_{i \in B}
  \varepsilon_i^{\id}(T,\muhat_i) - a\nB^2\,.
  \label{eq:evdW}
\end{equation}
The physical meaning of this pressure expression should be transparent
from the density, that is, the $\muB$-derivative of the pressure:
\begin{equation}
  \nB(T,\bmu) = f(\eta_B) \sum_{i \in B} n_i^{\id}(T,\muhat_i)\,.
  \label{eq:vdwhrg1}
\end{equation}

For the conventional VDW model, the standard choice of $f(\eta)$ is
\begin{equation}
  f_{\text{VDW}}(\eta) = 1-4\eta\,.
  \label{eq:fvdW}
\end{equation}
By definition the range of $f_{\text{VDW}}(\eta)$ must be limited
within $0 < f_{\text{VDW}}(\eta) \le 1$, where $f_{\text{VDW}}=1$
refers to the no EV limit and $f_{\text{VDW}}=0$ corresponds to the maximal packing.
When the maximal packing is reached at $\eta=1/4$,
the thermodynamic quantities become singular.
Singularity in the thermodynamic quantities at the maximal packing
leads to the acausal behavior in the speed of sound $c_s > 1$, where
the speed of sound is defined as $c_s^2 \equiv \partial p / \partial
\varepsilon$.
In the neutron star environment, using the hard-sphere radius
$R_B=0.511\fm$ (which will be determined in Sec.~\ref{sec:ns}), we
have a rough estimate for the maximal packing density as to $\nB
\simeq 2.8\,n_0$ and the density at which the superluminal speed of
sound ($c_s>1$) is reached as to $\nB \simeq 2.2\,n_0$;
$n_0=0.16\fm^{-3}$ is the normal nuclear density.
These values are lower than the density in central cores of
typical-mass neutron stars, and the above simple
choice~\eqref{eq:fvdW} is obviously inappropriate.

\subsection{Carnahan-Starling (CS) refinement of the excluded volume term}
\label{sec:csev}

In the VDW-type EV treatment, the interaction sphere is
too rigid and the maximal packing occurs at unphysically low density.
It would be sensible to smear the interaction clouds so that the EV
effects can mildly set in.
For the astrophysical application, therefore, we should improve the
function~\eqref{eq:fvdW} and a promising candidate is the
Carnahan-Starling (CS)-type EV~\cite{Carnahan:1969}.

For the CS-HRG model~\cite{Vovchenko:2017cbu}, the choice of $f(\eta)$ is
\begin{equation}
  f_{\text{CS}}(\eta) = \exp\left[ -\frac{(4-3\eta)\eta}{(1-\eta)^2} \right]\,.
  \label{eq:fCS}
\end{equation}
When $\eta\ll1$, we can easily confirm that
$f_{\text{CS}}(\eta) \approx 1 - 4\eta =
f_{\text{VDW}}(\eta)$ up to the linear order in $\eta$.
Here again, the range of $f_{\text{CS}}$ must be
limited within $0 < f_{\text{CS}} \le 1$.  The maximal packing occurs
at $\eta = 1$.  The corresponding density, $\nB=4\eta/b=3/(4\pi R_B^3)$,
then reaches $\nB=11.2\,n_0$ with $R_B =0.511\fm$.
Because of a longer tail up to $\eta\sim 1/2$ in Eq.~\eqref{eq:fCS} as
compared to a sharp drop at $\eta=1/4$ in Eq.~\eqref{eq:fvdW},
the maximal packing density in the CS-type EV is about four times
larger than that of the VDW-type EV, which extends the validity range.
This is a na\"{i}ve estimate, and the serious calculation concerning
the causality gives a rather smaller value of the limiting density, i.e.,
$\nB \simeq 3.7\,n_0$ for $R_B = 0.511\fm$.

\subsection{Need for the attractive interaction}

In the spirit of the HRG model, the attractive interaction leading to the
resonance formation is implicitly incorporated through added
resonances; however, not all the
attractive interactions are taken into account by resonances.
The nuclear force has attractive regions, which are
dominated by one-pion exchange, as well as repulsive regions by the exchange of heavy
mesons and multi-pions, as seen also in the first-principles lattice-QCD calculation~\cite{Ishii:2006ec}.
The VDW model is not the direct description of the nuclear
  force itself but it can emulate such characteristics of the nuclear
force;  the EV effect captures the repulsive core nature at short
range, while the intermediate and long-range parts are reasonably
captured by resonances and the attractive interaction term in the VDW model.

One might wonder how that the attractive term $\propto a$ affects
physical observables.  Only to avoid singular behavior of
thermodynamic quantities, the minimal model with either VDW- or
CS-type EV would be enough.  We would, however, stress that the
attractive term is indispensable for quantitative analysis.
We will come back to this point when we discuss the $M$-$R$ relation
using the EoS from the CS-HRG model later.

\section{EoS construction}
\label{sec:eos}

\begin{figure}
  \centering
  \includegraphics[width=.99\columnwidth]{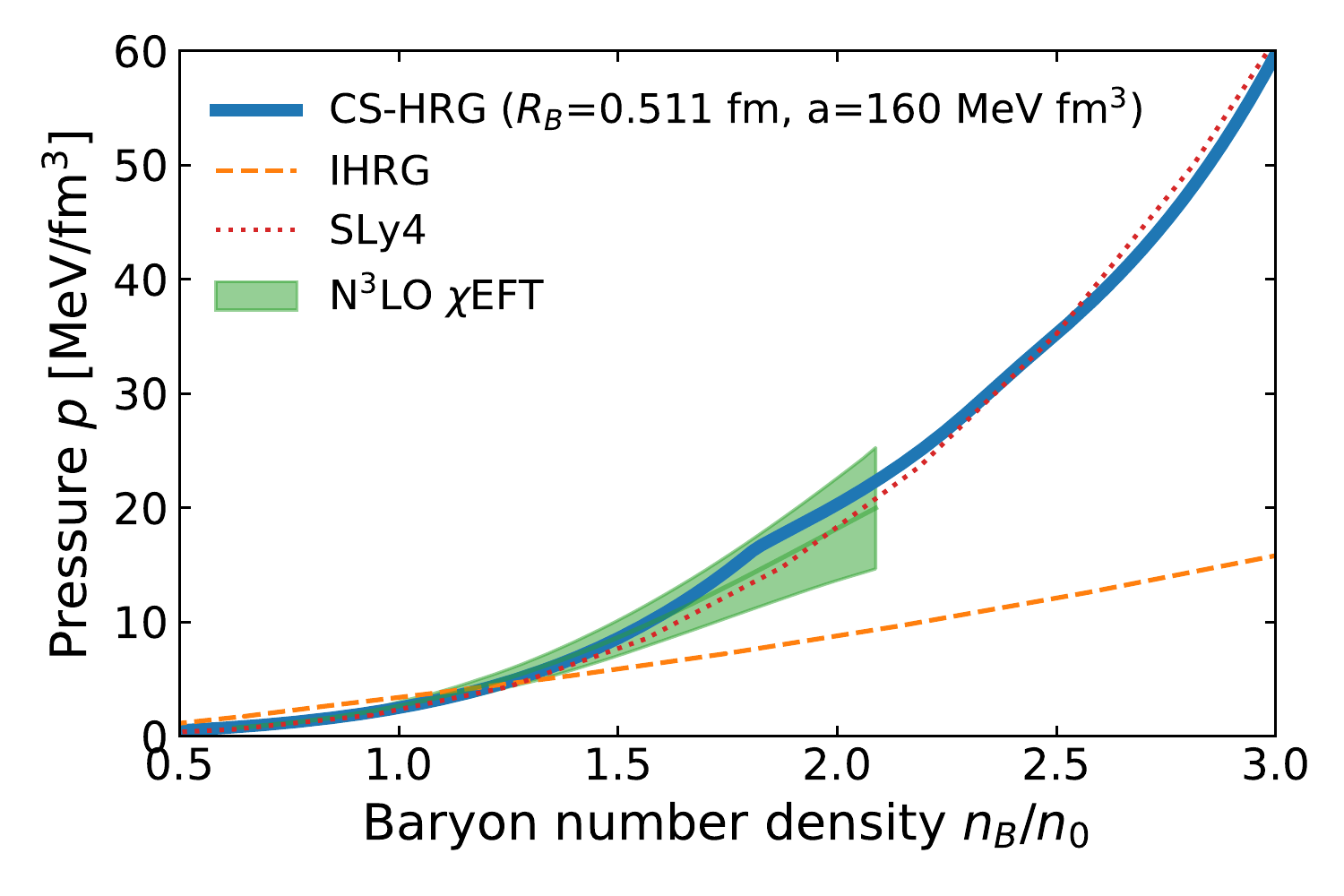}
  \caption{EoS for neutron star matter calculated from the CS-HRG model at $T=0$.
    The IHRG results and the phenomenological nuclear EoS
    (SLy4~\cite{Douchin:2001sv}), together with the N$^3$LO
    $\chi$EFT EoS~\cite{Drischler:2020fvz, Drischler:2020hwi} (to
    which our parameters are fitted) are shown.}
  \label{fig:zeroT_pnB}
\end{figure}

\begin{figure}
  \centering 
  \includegraphics[width=.99\columnwidth]{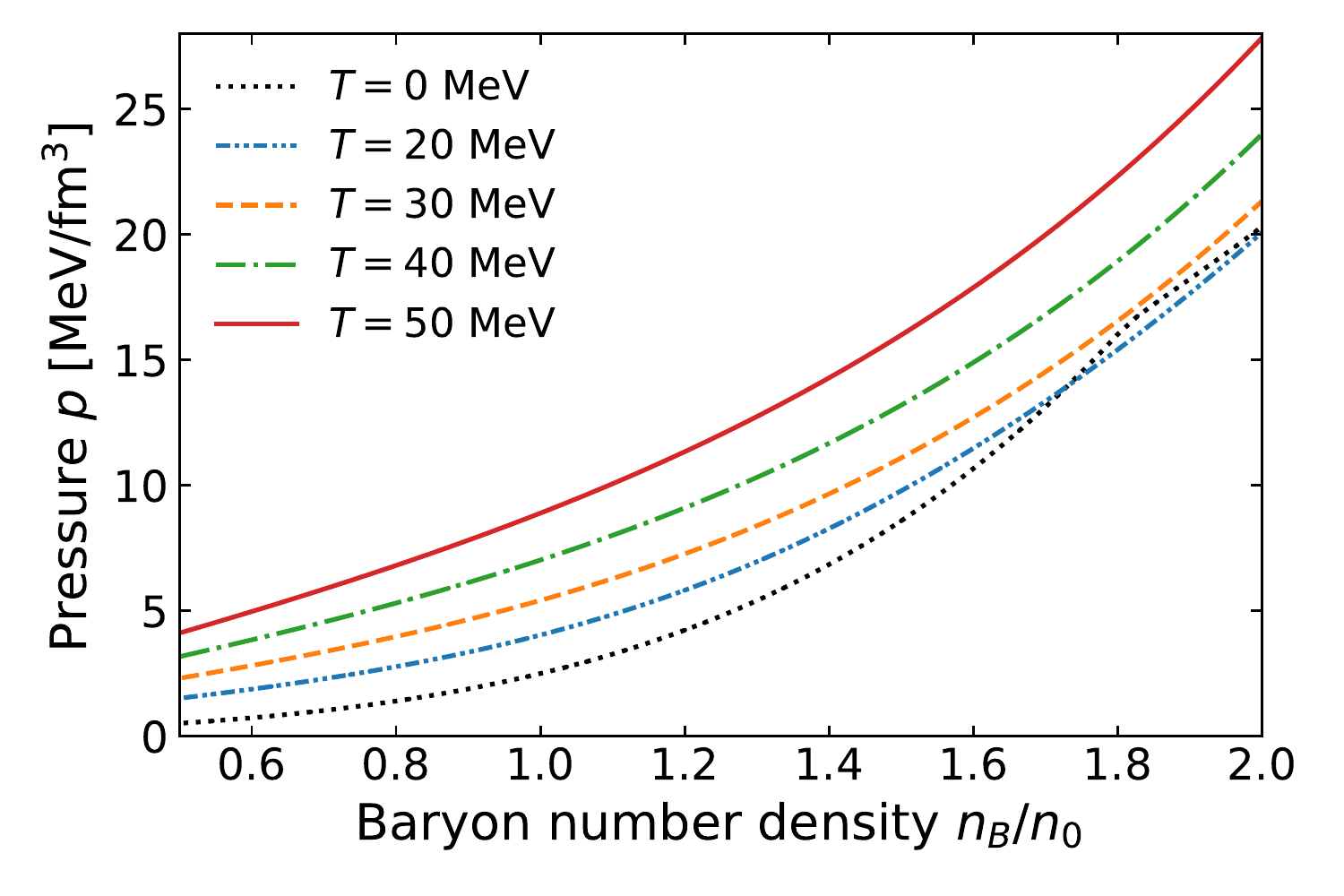}
  \caption{EoS calculated from the CS-HRG model at finite $T$.  The
    $T=0$ results are the same as shown in Fig.~\ref{fig:zeroT_pnB}
    but the scale is magnified.}
  \label{fig:finiteT_pnB}
\end{figure}

\begin{figure}
  \centering 
  \includegraphics[width=.99\columnwidth]{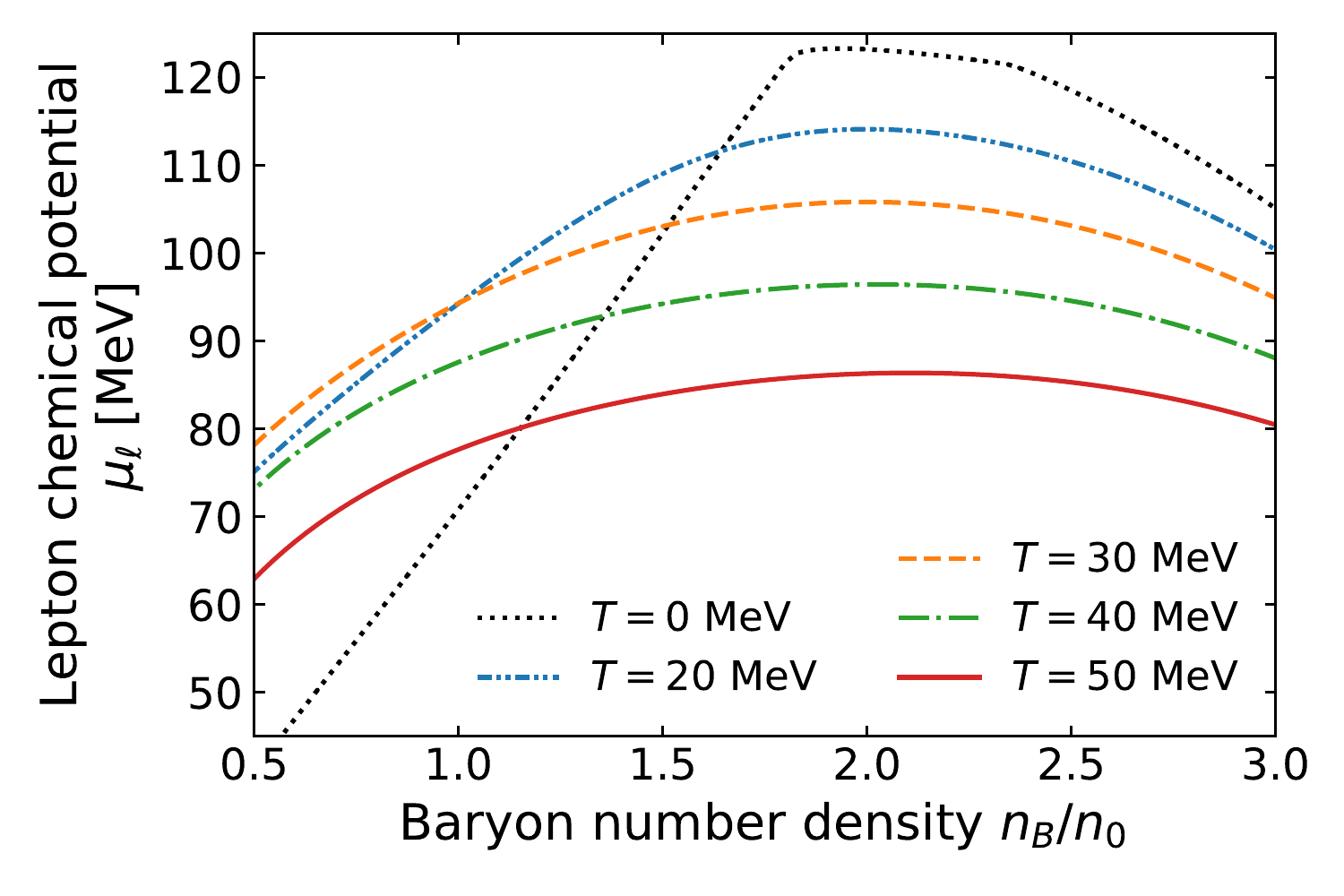}
  \caption{Lepton chemical potential $\mu_\ell = -\mu_Q$ obtained 
    by solving the electric charge neutrality and the 
    $\beta$-equilibrium conditions.}
  \label{fig:finiteT_mue}
\end{figure}

We are ready to construct the EoS for neutron stars at $T=0$ as
well as for compact-binary mergers and supernovae at $T>0$.  Under
these circumstances, the system is $\beta$-equilibrated via the weak
processes such as
\begin{align}
  n&\to p + \ell +\bar{\nu}_\ell\,, & p + \ell &\to n + \nu_\ell\,,\\
  \Sigma^-&\to n + \ell +\bar{\nu}_\ell\,, & n + \ell &\to \Sigma^- + \nu_\ell\,, 
\end{align}
and neutral in the electric charge.  These conditions of
$\beta$-equilibrium and electric charge neutrality amount
to setting the values of the chemical potentials, $\mu_Q$ and $\mu_S$.

Owing to large asymmetry of baryons and anti-baryons at high $\muB$,
we can safely neglect the anti-baryonic contribution to the EoS, and
we can also drop the mesonic contribution at $T=0$.
The meson condensation would cause subtlety; in particular, negatively
charged pions would form a condensate when $\mu_\ell > m_{\pi^-}$,
where $\mu_\ell$ is the leptochemical potential and
$m_\pi = 140\MeV$ is the pion mass.
We assume no condensate and this will be justified self-consistently
as shown in Fig.~\ref{fig:finiteT_mue}.
The values of $\mu_\ell$ at various temperatures in
Fig.~\ref{fig:finiteT_mue} do not exceed $m_\pi$.

We impose the following conditions on $\mu_Q$ and $\mu_S$:
\begin{align}
  & n_Q(T, \bmu) - n_\ell(T, \mu_\ell) = 0\,,\label{eq:neut}\\
  & \mu_S= 0\,,\label{eq:beta}
\end{align}
where $n_Q$ is the electric charge density in the interacting HRG model,
whose expression is given by
\begin{equation}
  \begin{split}
  \label{eq:nQ}
  n_Q(T, \bmu) =& f(\eta_B) \sum_{i\in B} Q_i n_i^\id (T, \muB + \Delta\muB + Q_i\mu_Q)\\
  &+ \sum_{i \in M} Q_i n_i^\id(T, Q_i\mu_Q)\,.
  \end{split}
\end{equation}
For the expression of the lepton density, $n_\ell$, we
substitute the electron mass, $m_e=0.511\MeV$, and the muon mass,
$m_\mu=106\MeV$ for the mass in the free particle expression~\eqref{eq:nGC}:
\begin{equation}
  \label{eq:nell}
  n_\ell(T, \mu_\ell) = n_{i = e}^{\id}(T, \mu_\ell; m_e) + n_{i = \mu}^{\id}(T, \mu_\ell; m_\mu)\,.
\end{equation}
We neglected the neutrino contribution by assuming that neutrinos
quickly escape in neutron star systems both for $T=0$ and $T>0$, which
is justified for ordinary neutron stars.
Now we can solve these conditions to find
\begin{equation}
  \mu_Q + \mu_\ell = 0\,,\quad \mu_S = 0\,.
\end{equation}
For the actual procedures, we fix the value of $\nB$, and solve
Eqs.~(\ref{eq:vdwhrg1})-(\ref{eq:vdwhrg3}), (\ref{eq:neut}), and (\ref{eq:beta})
in terms of three variables, $\muB$, $\mu_Q (=-\mu_\ell)$, and $\Delta \mu_B$.
Once three variables are obtained, we can compute the
thermodynamic quantities correspondingly.
We note in passing that $\muB$ and $\Delta\muB$ always appear in the
combination of $\hat\mu_B \equiv \mu_B + \Delta\mu_B$, so that
we can separately treat Eq.~\eqref{eq:vdwhrg3} and solve two coupled
equations to determine $\muhat_B$
and $\mu_Q$.

With bearing the application to neutron stars in mind, we calculate 
the EoS at $T=0$ following the procedures outlined above.
We plot our main results for the EoS from the CS-HRG model in
Fig.~\ref{fig:zeroT_pnB}.
The two parameters, $R_B$ and $a$, are fitted to reproduce the N$^3$LO 
$\chi$EFT EoS, which is taken from Ref.~\cite{Drischler:2020fvz} based 
on the pure neutron and symmetric nuclear matter 
calculations~\cite{Drischler:2020hwi}. 
It is evident from the comparison between our results (by the thick curve)
and the IHRG results (by the dashed curve) that the interaction is
crucial for neutron star matter.  We also observe that our fitted EoS
stays close to the phenomenological nuclear EoS, for which we choose
SLy4~\cite{Douchin:2001sv}.  It should be noted that the VDW-HRG model
with $f_{\text{VDW}}(\eta)$ in Eq.~\eqref{eq:fvdW} can by no means fit the
N$^3$LO $\chi$EFT nor the SLy4 EoS at all.  In
Fig.~\ref{fig:zeroT_pnB} there appears a bump around the $n_B =
1.8\,n_0$; it is related to the onset of hyperons as discussed
later in Fig.~\ref{fig:fraction}.  Here, too rapid stiffening is tamed
by softening induced by the appearance of hyperons.

The interacting HRG model has originally been used in the finite-$T$
circumstances, so it should be a reasonable setup for generalizing the
EoS construction to the finite-$T$ case.  In this way, the finite-$T$
EoS has been obtained from the CS-HRG model and we plot the results at
$T=0$, $20$, $30$, $40$, and $50\MeV$ in Fig.~\ref{fig:finiteT_pnB}.
Overall, the pressure becomes higher with increasing $T$ as expected, 
but we find an exception;  the EoS at $T=20\MeV$ goes below the $T=0$
EoS around $\nB\simeq 1.8\,n_0$.  We can understand this as follows. 
At $T=0$ new particles cannot appear until the density exceeds the 
mass threshold.  Below the onset density, the EoS becomes stiffer as 
the density rises up.  On the other hand, at finite $T$, the EoS can 
accommodate any massive particles according to the thermal weights, 
which makes a qualitative difference from the $T=0$ EoS and explains 
the change of the ordering around $\nB\simeq 1.8\,n_0$.

Finally, we show the leptochemical potential $\mu_\ell$ in
Fig.~\ref{fig:finiteT_mue}.  As we already mentioned before,
$\mu_\ell$ never exceeds the pion mass for any temperature and
density.  This self-consistently justifies our assumption of no meson
condensate.

\section{Astrophysical applications}
\label{sec:astro}

We will showcase the numerical results for the EoS and the
phenomenological implications at $T=0$ and finite $T$ in order.
The density and temperature reached in supernovae are as high as $\nB
\sim 2\mbox{-}3\,n_0$ and $T\sim 100\MeV$ while those in neutron star
mergers are $\nB \gtrsim 5\,n_0$ and $T\lesssim 50\MeV$.

\subsection{Neutron star matter and structure at $T=0$}
\label{sec:ns}

\begin{figure}[t]
  \centering 
  \includegraphics[width=.99\columnwidth]{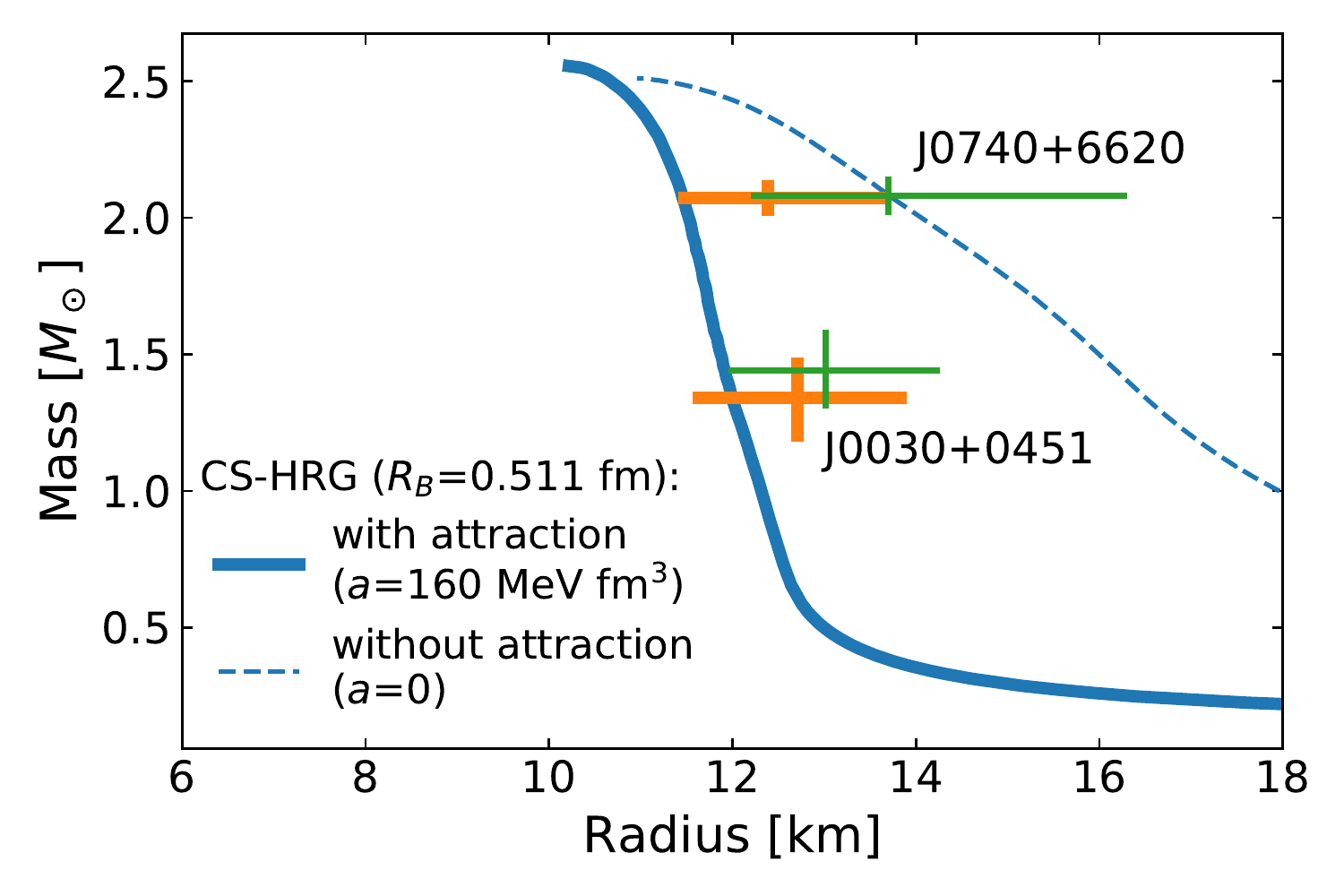}
  \caption{$M$-$R$ relation corresponding to the EoS in 
    Fig.~\ref{fig:zeroT_pnB}.  The thick (dashed) curve represents the
    result with (without) the attractive interaction. 
    The radius constraints from the NICER collaboration are 
    shown:
    (J0030+0451 at $M=1.4 \, M_\odot$) Ref.~\cite{Riley:2019yda} and 
    Ref.~\cite{Miller:2019cac} conclude a smaller (84~\% CL) 
    and a larger (68~\% CL) radius, respectively. 
    (J0740+6620 at $ M = 2\,M_\odot$) Ref.~\cite{Riley:2021pdl} and 
    Ref.~\cite{Miller:2021qha} conclude a smaller (84~\% CL) and 
    a larger (68~\% CL) radius, respectively.}
  \label{fig:zeroT_MR}
\end{figure}

\begin{figure}[t]
  \centering 
  \includegraphics[width=.99\columnwidth]{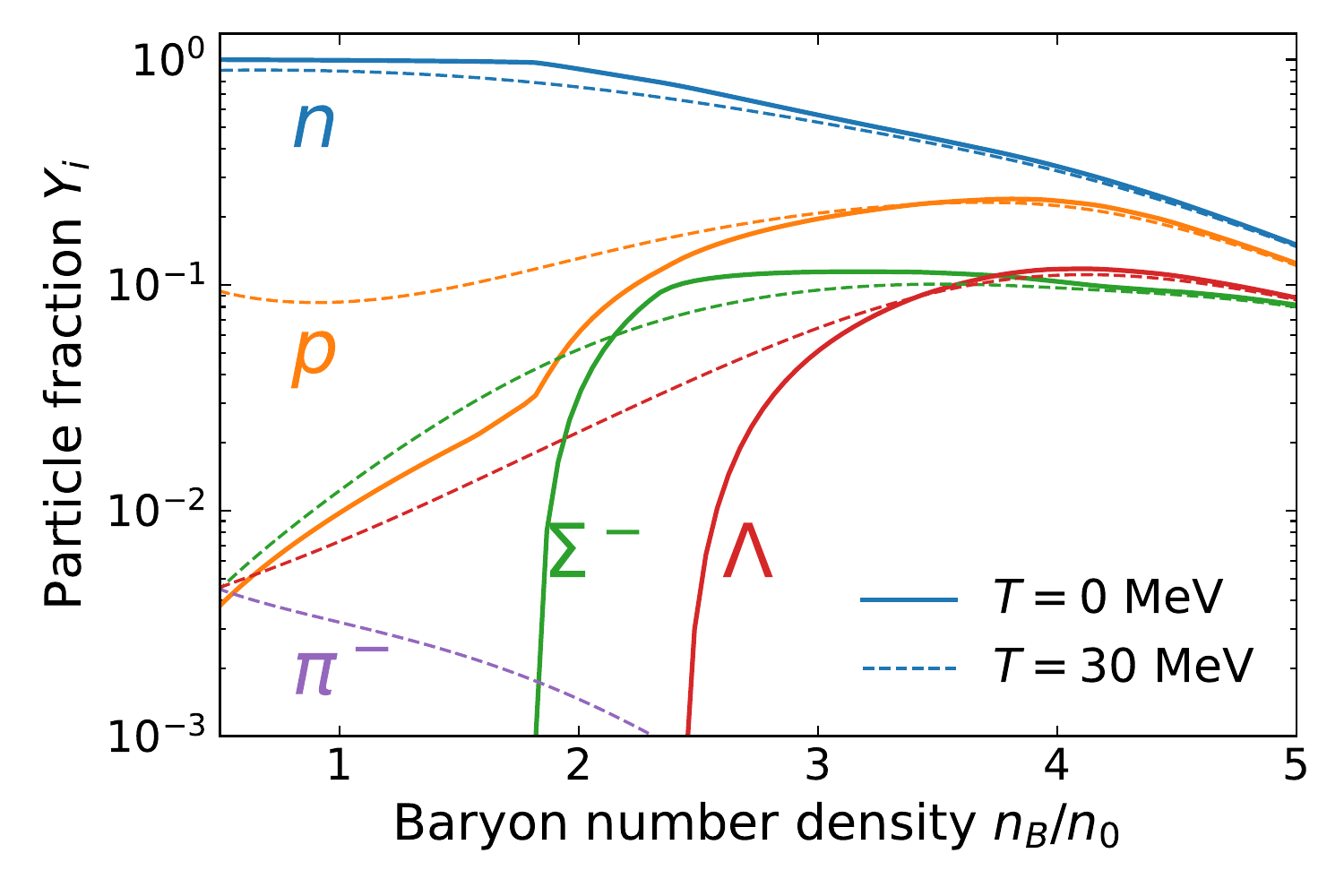}
  \caption{Particle fraction $Y_i = n_i / \nB$ corresponding to the
    EoS in Fig.~\ref{fig:zeroT_pnB}.  The finite-$T$ results are
    overlaid by dotted curves.}
  \label{fig:fraction}
\end{figure}

With the EoS from the CS-HRG model, we calculate the mass-radius
($M$-$R$) relation by solving the Tolman-Oppenheimer-Volkoff (TOV)
equation.  As shown in Fig.~\ref{fig:zeroT_MR},
our EoS incidentally matches with the phenomenologically accepted EoSs
such as SLy4, even in the crust region of neutron stars, and so we use
our EoS down to the surface of the star when solving the TOV equation;
see Ref.~\cite{Weinberg:1972kfs}.
We compare the cases with/without the attractive interaction.
If we turn off the attractive interaction at $a=0$, the radius at
$M = 1.4\,M_\odot$ is too large and out of the NICER observation as seen
in Fig.~\ref{fig:zeroT_MR}.
Usually, the lower density part of the EoS is responsible for the
radius of stars, and a larger radius is favored for a stiffer EoS at
low density.
This tendency can be confirmed in Fig.~\ref{fig:zeroT_pnB};
the IHRG EoS is constructed without the attractive interaction, and the
lower density part ($\nB/n_0 \lesssim 1$) from the IHRG model exceeds
that from the CS-HRG model.
This behavior is consistent with such an interpretation that the EoS
without attractive interaction gives too stiff EoS at low density and
leads to a too large radius.

As mentioned earlier, for $R_B = 0.511\fm$, the maximum packing
density, above which the model breaks down, is $n_B = 11.2\,n_0$ for
the CS-HRG model.
On the $M$-$R$ relation, the maximum mass of $M=2.56\,M_\odot$ is
attained at $\nB = 9.32\,n_0$, which certainly lies within the
validity range of the CS-HRG model.
Regarding the maximum mass, some controversies are unavoidable.
Combining the GW170817 event with the accompanying electromagnetic
observation, the maximum mass could be constrained to be at most
$\lesssim 2.3\,M_\odot$ ~\cite{Margalit:2017dij, Shibata:2017xdx,
  Rezzolla:2017aly, Ruiz:2017due}.
Meanwhile, a compact object with $\sim 2.6\,M_\odot$ has been observed
in the GW190814, which may be identified as a massive neutron star.
Near the maximum mass region, another subtlety arises from a
possible transition to quark matter~\cite{Annala:2019puf, Fujimoto:2020tjc}.
It is a nontrivial question where the validity bound of our model
should be.
If we locate the validity bound at $\nB\simeq 3.7\,n_0$ (see
explanations in Sec.~\ref{sec:csev}), our model should be very apt up
to $M\simeq 1.5\,M_\odot$.

The HRG model provides us with a convenient picture to probe the
particle abundances.  In Fig.~\ref{fig:fraction} we show the
fraction $Y_i = n_i/\nB$ of the particle species $i$.
At small density, the neutron, $n$, is dominant with a small fraction
of the proton, $p$, which slowly increases with increasing density.
The onset of the hyperons is observed slightly below $2\,n_0$.  As is
consistent with the conventional scenario (see, e.g., Sec.~5 of
Ref.~\cite{Burgio:2021vgk}), $\Sigma^-$ is activated first as we
increase the density, and then $\Lambda$ is produced afterwards.

\subsection{Thermal index}
\label{sec:sn}


\begin{figure}
  \centering 
\includegraphics[width=.99\columnwidth]{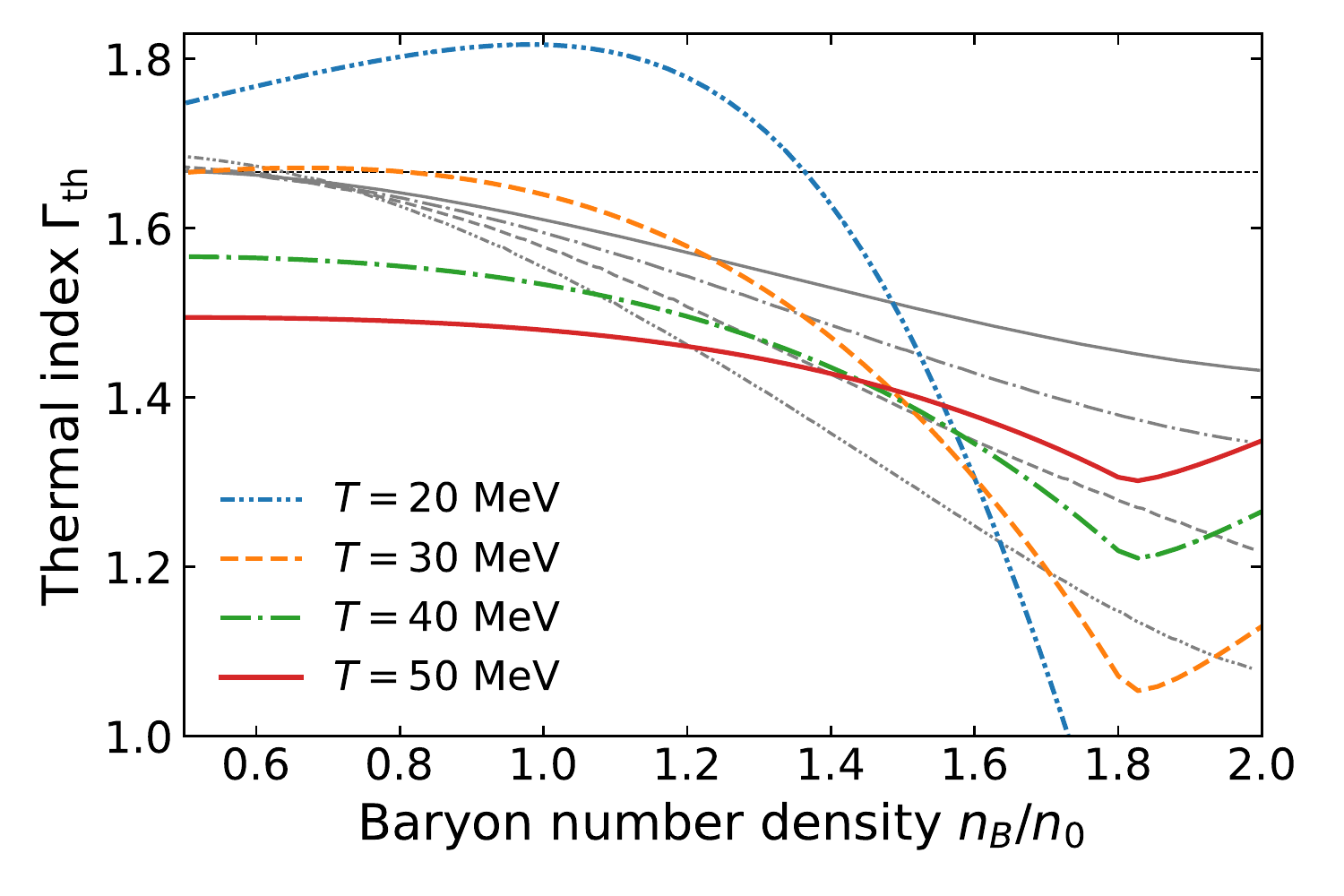}  
  \caption{Thermal index $\Gamma_\thermal$ corresponding to the EoS 
    shown in Fig.~\ref{fig:finiteT_pnB}.  For reference, the \textit{ab
      initio} calculations for pure neutron
    matter~\cite{Carbone:2019pkr} are overlaid in the grey color with
    the same line style as our results.  The canonical
    value of the adiabatic index for non-relativistic ideal gas,
    $\Gamma=5/3$, is shown by a horizontal line.}
  \label{fig:finiteT_Gammath}
\end{figure}

In the applications for astrophysical phenomena such as supernovae and
binary neutron star mergers, the thermal corrections to the EoS are
often modeled by an ideal gas
approximation~\cite{1993A&A...268..360J, Bauswein:2010dn,Figura:2020fkj}.
In order to define the thermal part of the EoS, which is parametrized
by the thermal index, $\Gamma_\thermal$, we introduce the rest-mass
density of baryons as $\rho_B = m_B \nB$ with the nucleon mass,
$m_B = 939\MeV$.
We can decompose the energy density $\varepsilon$ as
$\varepsilon =   (1+e)\rho_B$, where $e$ is the specific internal energy.
We can add the thermal corrections to the pressure and the energy on
top of the $T=0$ parts as
\begin{equation}
  p = p_{T=0} + p_{\thermal}\,,\qquad
  e = e_{T=0} + e_\thermal\,.
\end{equation}
In the simulations of neutron star mergers, the cold ($T=0$) component
is used before shock heating associated with the stellar collision
sets in.
The relation between the thermal pressure and the energy should be
supplemented with the additional constraint, that is commonly
parametrized by
\begin{equation}
  p_{\thermal}= (\Gamma_{\thermal}-1) \rho_{B} e_{\thermal}\,.
\end{equation}

In the phenomenological studies, $\Gamma_{\thermal}$ is a free
parameter.  It is customary to choose $\Gamma_{\thermal}$ around
$\sim 1.7$.  For example, $\Gamma_\thermal = 1.8$ was adopted in  
Ref.~\cite{Hotokezaka:2013iia}.
If $\Gamma_{\thermal}$ is too small (like $\sim 1.3$),
the thermal pressure is not large enough to sustain matter, resulting
in a rapid proto-neutron star contraction for supernovae, and in a
faster collapse of merger remnants to black holes for binary merger. 
In this way, smaller values of $\Gamma_{\thermal}$ may have impact on
core-collapse supernova and binary neutron star merger
simulations~\cite{Yasin:2018ckc}.
In contrast, a larger $\Gamma_{\thermal}$ (like $\sim 2.0$) would elongate the
life-time of the post-merger dynamics.
Thus, for reliable theoretical predictions, it is of utmost
importance to constrain $\Gamma_{\thermal}$.  Moreover,
although it is often assumed to be constant,
$\Gamma_{\thermal}$ may depend on the density and
temperature~\cite{Carbone:2019pkr}.

We can make use of our EoS to infer $\Gamma_{\thermal}$, which can be
represented in terms of the thermodynamic quantities as
\begin{equation}
  \Gamma_{\thermal} = 1 + \frac{p_{\thermal}}{e_{\thermal}\rho_{B}} 
    = 1 + \frac{p_{\thermal}}{\varepsilon_{\thermal}}\,.
\end{equation}
Here, $\varepsilon_{\thermal}=\rho_{B} e_{\thermal}$ is the thermal part of the energy density.
In Fig.~\ref{fig:finiteT_Gammath}, we show our estimate for the
thermal index, $\Gamma_\thermal$, as a function of the density.
We find that $\Gamma_\thermal$ becomes less sensitive to the density
as $T$ gets larger; e.g., $\Gamma_\thermal$ is almost constant around
$\sim 1.4$ at $T=50\MeV$.
The preceding \textit{ab initio} calculation based on
$\chi$EFT~\cite{Carbone:2019pkr} is also overlaid on
Fig.~\ref{fig:finiteT_Gammath}.
Our $\Gamma_\thermal$ and the \textit{ab initio}
$\Gamma_\thermal$~\cite{Carbone:2019pkr} differ qualitatively;  it
may be partially because the slope of the \textit{ab initio} EoS at
$T=0$ is gentle compared with the state-of-the-art $\chi$EFT
EoS~\cite{Drischler:2020fvz, Drischler:2020hwi} to which our model is
fitted to.
We note that $\Gamma_{\thermal}$ in Fig.~\ref{fig:finiteT_Gammath} is
computed under the assumption of neutrinoless $\beta$-equilibrium.
The values of $\Gamma_{\thermal}$ shown in this figure may deviate
from effective values realized in dynamical astrophysical phenomena,
particularly when neutrinos are trapped inside hot neutron stars.

\section{Summary and outlooks}
\label{sec:sum}

We demonstrated a successful construction of the equation of state
based on the van der Waals prescription of the hadron resonance gas
model with the Carnahan-Starling refinement of the excluded volume
term.
The ideal hadron resonance gas description does not work at high
density as shown in Fig.~\ref{fig:zeroT_pnB};  at
$T=0$ both the repulsive and the attractive interactions are crucial.
In our proposal a very simple parametrization can fit in with the empirical EoSs.
This provides us with an intuitive picture of the quarkyonic regime as
well as a paractically useful tool for phenomenological studies.
It is then a straightforward extension to include the finite-$T$
corrections.  We quantified the thermal index, $\Gamma_\thermal$, as a
function of the density for various temperatures.  Our estimated
$\Gamma_\thermal$ favors a rather smaller value than the
conventionally adopted one.

Our present modeling is the simplest one, and we can consider
improvements in many respects.  For example, the interaction should be
species dependent, and in particular, the strangeness sector could
behave differently.  In some channels involving the strangeness, the
Pauli blocking is relaxed and the short-range repulsive core is
absent.
As neutron stars contain hyperons as inevitable physical degrees of
freedom, the flavor-dependent interactions should be considered along
the line of Ref.~\cite{Vovchenko:2017zpj}.
It is also an interesting question to think of meson interaction
effects.
Moreover, since the validity region of our proposed model should cover
the high-$T$ and low-$\muB$ regime, we can apply our predicted EoS for
the heavy-ion collision experiments at lower collision energies.
These await to be investigated as future extensions.

\section*{Acknowledgments}
YF thanks Koichi~Murase for the consultation on numerical codes.
The authors thank
Koutarou~Kyutoku,
Koichi~Murase, and
Hiroyuki~Tajima
for useful discussions.
This work was supported by Japan Society for the Promotion of Science
(JSPS) KAKENHI Grant Nos.\ 18H01211, 19K21874, and 
20J10506.

\bibliography{hrgeos}
\bibliographystyle{elsarticle-num}

\end{document}